\input{aipcheck}

\documentclass[
    ,final            
  ]
  {aipproc}

\layoutstyle{6x9}

\begin{document}

\title{Signals of Warped Extra Dimensions at the LHC}

\classification{12.60.-i,11.10.Kk, 12.60.Cn}
\keywords{LHC, graviton, extra dimensions, spin}

\author{P.\ Osland}{
  address={Department of Physics and Technology, University of
  Bergen, N-5020 Bergen, Norway}}

\author{A.\ A.\ Pankov}{
  address={The Abdus Salam ICTP Affiliated Centre, Technical
  University of Gomel, 246746 Gomel, Belarus}}

\author{A.\ V.\ Tsytrinov}{
  address={The Abdus Salam ICTP Affiliated Centre, Technical
  University of Gomel, 246746 Gomel, Belarus}}

\author{N.\ Paver}{
address={University of Trieste and INFN-Trieste Section, 34100
Trieste, Italy}}

\begin{abstract}
We discuss the signatures of the spin-2 graviton excitations
predicted by the Randall-Sundrum model with one warped
extra dimension, in dilepton and diphoton production at LHC. By
using a specific angular analysis, we assess the ranges in mass
and coupling constant where such gravitons can be discriminated
against competitor spin-1 and spin-0 objects, that potentially
could manifest themselves in these processes with the same mass and
rate of events. Depending on the value of the coupling constant to
quarks and leptons, the numerical results indicate graviton
identification mass ranges up to 1.1--2.4~TeV and
1.6--3.2~TeV for LHC nominal energy of 14~TeV and time-integrated
luminosity of 10 and 100~${\rm fb}^{-1}$, respectively.
\end{abstract}

\maketitle


\section{INTRODUCTION}
Originally, extra spatial dimensions were proposed to
address the mass scale hierarchies affecting
the Standard Model (SM), and requiring parameter
fine tuning, in particular the so-called
gauge hierarchy
$M_{\rm EW}\ll M_{\rm Pl}\sim 10^{16}$~TeV. These
scenarios predict the existence of heavy new
particles, or excitations of the SM particles,
that could be revealed as narrow peaks in cross sections
measured at the LHC if their masses were within the
experimental kinematical reach. Observability will
depend on the values of these heavy `resonance'
masses and coupling constants to the SM matter, that
assure sufficiently  high event rates.
\par
Actually, in this regard, the {\it discovery reach} on an
individual scenario is the maximum value of the corresponding
resonance mass, $M_R$, for which the peak can be observed.
However, different non-standard models can give, for appropriate
values of their parameters, peaks at the same $M=M_R$ and the same
number of events under the peak. Accordingly, one defines for any
non-standard scenario an {\it identification} mass range (of
course, included within the discovery range), where it can be
unambiguously discriminated as the one really underlying the peak,
by excluding the other models as potential sources of that same peak.
The determination of the spin of a discovered heavy resonance is,
therefore, crucial for its identification. We will here discuss
the identification of the spin-2 graviton excitation predicted by
the Randall-Sundrum (RS) model with one warped extra
dimension \cite{Randall:1999ee}, against the spin-1 and the spin-0
hypotheses for a heavy neutral resonance discovered in dilepton
and diphoton inclusive production at the LHC:
\begin{equation}
p+p\to l^+l^-+X\ \ (l=e,\mu)\ \ \ {\rm and}
\ \ \ p+p\to\gamma\gamma+X.
\label{proc}
\end{equation}
We will model the spin-1 hypothesis by the $Z^\prime$s
predicted by extended electroweak gauge symmetries
\cite{Langacker:2008yv}, and the spin-0 one by the
sneutrinos ($\tilde\nu$) envisaged by $R$-parity violating
SUSY extensions of the SM \cite{Kalinowski:1997bc}.
`Confusion' domains of their respective parameter spaces
allowed by current experimental constraints exist, in which
`$s$-channel' exchanges of the above mentioned particles
can in the process (\ref{proc}) produce narrow peaks in
the dilepton invariant mass with the same values of
mass and event rates.
Therefore, for the spin-2 hypothesis
discrimination against the two alternative ones, more
detailed information must be supplied, i.e., that 
embodied in the characteristic angular distributions 
of the different scenarios.

\section{BASIC OBSERVABLES AND ANGULAR ANALYSIS}
The basic observable for a heavy resonance discovery
at an invariant dilepton (or diphoton) mass $M=M_R$
(with in our case $R=G,Z^\prime, {\tilde\nu}$ denoting
graviton, $Z^\prime$ and sneutrino, respectively) is the
production cross section governing the rate of events
\begin{equation}
\sigma{(pp\to R)} \cdot {\rm BR}(R \to l^+l^-)
=\int_{-z_{\rm{cut}}}^{z_{\rm cut}}{\rm d} z
\int_{M_{R}-\Delta M/2}^{M_{R}+\Delta M/2}{\rm d} M
\int_{y_{\rm min}}^{y_{\rm max}}{\rm d} y
\frac{{\rm d}\sigma}{{\rm d} M\, {\rm d} y\, {\rm d} z},
\label{TotCr}
\end{equation}
and the differential angular distribution
\begin{equation}
\frac{{\rm d}\sigma}{{\rm d} z} =
\int_{M_{R}-\Delta M/2}^{M_{R}+\Delta M/2}{\rm d} M
\int_{y_{\rm min}}^{y_{\rm max}}
\frac{{\rm d}\sigma}{{\rm d} M\, {\rm d} y\, {\rm d} z}\,{\rm d} y.
\label{DiffCr}
\end{equation}
In Eqs.~(\ref{TotCr}) and (\ref{DiffCr}), $z=\cos\theta_{\rm cm}$
and $y$ define the lepton-quark (or photon-quark) angle in the
dilepton (or diphoton) center-of-mass and the dilepton rapidity,
respectively, and cuts on phase
space due to detector acceptance are indicated. For integration
over the full phase space, the limits would be $z_{\rm cut}=1$
and $y_{\rm max}=-y_{\rm min}=\log({\sqrt s}/M)$
with $\sqrt s$ the LHC collider center-of-mass energy.
Furthermore, $\Delta M$ is an invariant mass bin
around $M_R$, reflecting the detector energy resolution,
see for instance Ref.~\cite{Atlas}. To evaluate the number
$N_S$ of resonant signal events, these equations must be
multiplied by the time-integrated luminosity, for which we
will take 100 and 10 ${\rm fb}^{-1}$, and by the foreseen
reconstruction efficiencies (90\% for both electrons and
muons \cite{Cousins:2004jc}). Also, the typical experimental
cuts on rapidities and transverse momenta must be applied
($p_\bot >20$~GeV and $\vert\eta\vert < 2.5$ for both leptons).
Finally, with $N_B$ the number of `background'
events in the $\Delta M$ bin, determined by the SM predictions,
the criterion $N_S=5{\sqrt{N_B}}$ or 10~events, whichever is
larger, will be adopted as the minimum signal for the peak
discovery. Of course, the determination of discovery and
identification reaches on the different non-standard models
requires the expression of  Eqs.~(\ref{TotCr}) and
(\ref{DiffCr}) in terms of convolutions of the pertinent
partonic cross sections times parton distribution functions,
and for the latter we will choose the CTEQ6 ones of
Ref.~\cite{Pumplin:2002vw}.
\par
Since the $z$-dependence of the cross sections for
graviton, $Z^\prime$ and sneutrino exchanges are different,
the discrimination of spin-2 from spin-1 and spin-0
might be attempted by the `direct' comparison of
Eq.~(\ref{DiffCr}) for the three hypotheses
\cite{Allanach:2000nr,Cousins:2005pq}.
\par
In practice, due to the completely symmetric $pp$
initial state at the LHC, the determination of the sign of
$z$ from the measured events may not be fully unambiguous.
An observable potentially avoiding this ambiguity, and
which we adopt to perform the angular analysis, is the
$z$-evenly integrated center-edge angular asymmetry
\cite{Dvergsnes:2004tw,Osland:2008sy}:
\begin{equation}
\label{ace}
A_{\rm{CE}}=\frac{\sigma_{\rm{CE}}}{\sigma}\quad{\rm
with} \quad \sigma_{\rm{CE}} \equiv \left[\int_{-z^*}^{z^*} -
\left(\int_{-z_{\rm cut}}^{-z^*} +\int_{z^*}^{z_{\rm
cut}}\right)\right] \frac{{\rm d} \sigma}{{\rm d} z}\, {\rm d} z.
\end{equation}
In Eq.~(\ref{ace}), $0<z^*<z_{\rm cut}$ defines
the separation between the ``center'' and the ``edge''
angular regions, and is {\it a priori} arbitrary.
However, its actual value can be ``optimized'' in the
numerical analysis. A further
potential advantage of $A_{\rm CE}$ is that, as
consisting of ratios of integrated cross
sections, it could be less sensitive to systematic
uncertainties, such as those stemming from particular
choices of parton distributions, $K$-factor values,
etc..\footnote{Recently, an asymmetry
${\tilde A}_{\rm CE}$ defined along these lines, in terms
of differences between lepton and antilepton
pseudorapidities, has been proposed for model
identification in Ref.~\cite{Diener:2009ee}.
Also, the possibility of identifying
the graviton excitation spin-2 from
the azimuthal angular dependence of the graviton+jet
inclusive production has been explored
in \cite{Murayama:2009jz}.}

\section{NON-STANDARD MODELS AND ANGULAR DISTRIBUTIONS}
%
%
\leftline{\bf RS model with one compactified warped extra dimension}
This simplest version consists of one warped extra
dimension, $y$, two three-dimensional branes placed at a
compactification relative distance $\pi R_c$ in $y$,
and the specific 5-D metric \cite{Randall:1999ee}
\begin{equation}
\label{metric}
ds^2=\exp{(-2k\vert y\vert)}\ \eta_{\mu\nu}dx^\mu dx^\nu-dy^2,
\end{equation}
where $\eta_{\mu\nu}$ is the usual Minkowski tensor and $k>0$
is a dimensionful constant. The SM fields are assumed to be
localized to the so-called TeV brane, while gravity can propagate
in the full 5-D space, in particular on the other brane,
the Planck brane, in which the effective 4-D mass scale is
${\overline M}_{\rm Pl}=1/\sqrt{8\pi G_{\rm N}}=
2.44\times 10^{15}\, {\rm TeV}$. With $M_*$ the 5-D mass scale,
analogously related to the cubic root of the 5-D gravitational
constant, Einstein's equations imply the relation
${\overline M}_{\rm Pl}^2=({M_*}^3/k)(1-\exp{(-2k\pi R_c)})$, and
the basic `naturalness' assumption imposed on the model, to
avoid further fine tunings, is
${\overline M}_{\rm Pl}\sim M_*\sim k$. The geometry
of Eq.~(\ref{metric}) implies that the mass spectrum on
the Planck brane, of the order of $10^{15}$ TeV, can
for $kR_c\sim 11$ be exponentially `warped' down to the TeV
brane where SM particles live, by many orders of magnitude,
namely, to the scale
$\Lambda_\pi={\overline M}_{\rm Pl}\exp{(-k\pi R_c)}\sim 1$ TeV.
The appealing consequence is then that gravitational effects can
occur in the reach of supercolliders, such as the LHC. Indeed,
junction conditions at the brane $y$-positions
imply, for the above value of $kR_c$, the existence
of a tower of spin-2 graviton excitations, $h_{\mu\nu}^{(n)}$,
with a specifically spaced mass spectrum
$M_n=x_n k\exp{(-k\pi R_c)}$ of order TeV ($x_n$
are here the roots of $J_1(x_n)=0$). Their couplings
to the SM particles are only $1/\Lambda_\pi$-suppressed (not
$1/{\overline M}_{\rm Pl}$):
\begin{equation}
\label{interaction}
{\cal L}_{\rm TeV}=
-\left[\frac{1}{{\overline M}_{\rm Pl}}h_{\mu\nu}^{(0)}(x)+
\frac{1}{\Lambda_\pi}\sum_{n=1}^{\infty}h_{\mu\nu}^{(n)}(x)\right]
T^{\mu\nu}(x),
\end{equation}
and their signature may appear at LHC. In (\ref{interaction}),
$T^{\mu\nu}$ is the energy-momentum tensor and
$h_{\mu\nu}^{(0)}$ denotes the zero-mode, ordinary, graviton.
\par
The commonly chosen independent
RS model parameters are the mass of the lowest graviton
excitation, $M_G\equiv M_1$, and the `universal',
dimensionless,  coupling $c=k/{\overline M}_{\rm Pl}$
(the scale $\Lambda_\pi$ is then a derived
quantity). Theoretically `natural' ranges of these parameters
are $0.01\leq c\leq 0.1$ and
$\Lambda_\pi< 10$ TeV \cite{Davoudiasl:2000jd}. With
$\Gamma_n=\rho M_n x_n^2 c^2$ and $\rho$ a number of the order
of $0.1$, narrow graviton resonances are expected for
such small values of $c$.
\par
Tevatron 95\% CL limits on $M_G$ \cite{Abazov:2010xh}
range from $M_G>560$ GeV ($c=0.01$) to
$M_G>1.050$ TeV ($c=0.1$).
Thus, due to the high allowed values of $M_G$,
the discovery could be limited to the lightest $M_1$ and
the verification of the predicted mass pattern hardly
feasible, so that the spin-2 determination
through the angular analysis becomes a crucial test
of the model.
\par
For dilepton production, the leading order subprocesses
${\bar q}q\to G\to l^+l^-$ and $gg\to G\to l^+l^-$
give for the $z$-even distributions needed in $A_{\rm CE}$,
with self-explaining notations \cite{Han:1998sg,Giudice:1998ck}:
\begin{equation}
\frac{{\rm d}\sigma^G}{{\rm d}z}=
\frac{3}{8}(1+z^2)\sigma_{q}^{\rm SM} +
\frac{5}{8}(1-3z^2+4z^4)\sigma^G_{q} +
\frac{5}{8}(1-z^4)\sigma^G_{g},
\label{Diffg}
\end{equation}
and:
\begin{equation}
A_{\rm CE}^{G} =\epsilon_q^{\rm SM}\,A_{\rm CE}^{\rm SM} +
\epsilon^G_q\left[2\,{z^*}^5+\frac{5}{2}\,z^*(1-{z^*}^2)-1\right]
+ \epsilon^G_g\left[\frac{1}{2}\,{z^*}(5-{z^*}^4)-1\right].
\label{aceg}
\end{equation}
In (\ref{aceg}), $\epsilon^G_q$, $\epsilon^G_g$ and
$\epsilon_q^{\rm SM}$ are the fractions of $G$-events under the
peak at $M_R$ initiated by ${\bar q} q$ and $gg$ processes, and
the SM background, respectively. They are determined by overlaps
of parton distribution functions and, obviously,
$\epsilon^G_q+\epsilon^G_g+\epsilon_q^{\rm SM}=1$. Strictly,
Eqs.~(\ref{Diffg}) and (\ref{aceg}) are quite transparent in
showing the characteristic $z$ (and $z^*$) dependencies for the
spin-2 graviton, but hold in that form only for $z_{\rm cut}=1$.
This fact turns out to be numerically unimportant at the optimal
value of $z^*$ where the $A_{\rm CE}$ analysis is performed, and
the final results take into account all phase space cuts foreseen
by the experiment. Moreover, next-to-leading order terms in QCD
\cite{Mathews} have for simplicity been included in the
calculations through flat (in $z$) $K$-factors, $K\simeq 1.3$.
\par
For diphoton production (\ref{proc}), retaining
only the leading order RS resonance exchange
contributions to simplify the presentation,
${\bar q}q \to G \to \gamma\gamma$ and
$gg \to G \to \gamma\gamma$,
the analogues of Eqs.~(\ref{Diffg}) and
(\ref{aceg}) with the same significance of the
notations can be written as \cite{Sridhar:2001sf}:
\begin{equation}
\frac{{\rm d}\sigma^G}{{\rm d}z} =
\frac{5}{8} (1-z^4) \sigma_q^G +
\frac{5}{32} (1 + 6 z^{2} + z^{4}) \sigma_g^G,
\label{Diffggam}
\end{equation}
and
\begin{equation}
\label{aceggam}
A_{\rm CE}^{G} =
\epsilon_q^G \left[\frac{1}{2} z^{*} (5-{z^{*}}^{4}) - 1\right] +
\epsilon_g^G \left[-1 + \frac{5}{8}z^{*} + \frac{5}{4}{z^{*}}^{3} +
\frac{1}{8}{z^{*}}^{5}\right].
\end{equation}
It turns out that in this case the $gg$-initiated subprocess is
the dominant, with a shape peaked at $z=\pm 1$ similar to the SM
background but by far overwhelming it for the order TeV values of
$M_G$ we here are interested in. Clearly, the interest of the
diphoton channel is that spin-1$\not\to\gamma\gamma$ leaves only
the spin-2 and spin-0 hypotheses, and that the ratio ${\rm
BR}(G\to\gamma\gamma)/{\rm BR}(G\to l^+l^-)\simeq 2$ is predicted
\cite{Han:1998sg}.
\par
A current field of interest is the extension of
the simplest RS model described here to the case 
of SM fermions and
gauge bosons also propagating in the full 5-D space,
this could provide an approach also to other mass
hierarchies such as, for example, the fermion mass
hierarchy. A complicated phenomenology
then emerges, in which both fermions and gauge bosons
are accompanied by high (and very high) mass excitations,
see for instance the review in Ref.~\cite{Davoudiasl:2009cd}.
The existence of the spin-2 graviton excitations
remains, however, a general feature.\footnote{The
other one being represented by the spin-0,
massive, radion needed to stabilize the
compactification radius
$R_c$ \cite{Goldberger:1999uk}.}

%
%
\medskip
\par\noindent
{\bf Heavy neutral gauge bosons}
\par\noindent
Turning to the spin-1 hypothesis, $Z^\prime$ models
depend, besides the mass $M=M_{Z^\prime}$, on left- and
right-handed couplings to quarks and leptons. In popular
scenarios we refer to, generated by different extended
electroweak gauge symmetries, those couplings are fixed
theoretically, so that only $M_{Z^\prime}$ would remain
as a free parameter. This is the case of the
$Z^\prime_\chi$, $Z^\prime_\psi$, $Z^\prime_\eta$,
$Z^\prime_{\rm LR}$, $Z^\prime_{\rm ALR}$ models, and the
``sequential'' $Z^\prime_{\rm SSM}$ model with the same
couplings as the SM (details can be found
in \cite{Langacker:2008yv}). Current Tevatron lower
limits (95\% CL) on these $Z^\prime$ masses range from
878 GeV to 1.03 TeV, depending on the
model \cite{Aaltonen:2008ah}.
\par
Actually, the leading-order partonic subprocess
${\bar q}q\to {Z^\prime}\to l^+l^-$ leads to
the same form of the $z$-even angular distribution
as the SM, therefore to the same $A_{\rm CE}$ for all models:
\begin{equation}
\frac{{\rm d}\sigma^{Z^\prime}}{{\rm d}z}=\frac{3}{8}
(1+z^2)[\sigma_{q}^{\rm SM}+ \sigma^{Z^\prime}_{q}]
\label{Diffzprime}
\end{equation}
and
\begin{equation}
\label{acezprime}
A_{\rm CE}^{Z^\prime}\equiv A_{\rm CE}^{\rm SM}=
\frac{1}{2}z^*(z^{*2}+3)-1.
\end{equation}
Our $A_{\rm CE}$-based  estimates for exclusion of the
spin-1 $Z^\prime$ hypothesis will, accordingly, have
a considerable degree of $Z^\prime$ model-independence.
%
%
\medskip
\par\noindent
{\bf Sneutrino exchange}
\par\noindent
The spin-0 character of the resonant subprocess
${\bar d}d \to{\tilde\nu}\to l^+l^-$, leading to a peak
at $M=M_{\tilde\nu}$, manifests itself in a flat angular
distribution \cite{Kalinowski:1997bc}:
\begin{equation}
\frac{{\rm d}\sigma^{\tilde\nu}}{{\rm d} z}=
\frac{3}{8}(1+z^2)\sigma^{\rm SM}_{q}+ \frac{1}{2}\sigma
^{\tilde\nu}_{q},
\label{Diffsneu}
\end{equation}
\begin{equation}
A_{\rm CE}^{\tilde\nu} = \epsilon_q^{\rm SM}\,A_{\rm CE}^{\rm SM}
+\epsilon^{\tilde\nu}_{q}(2z^*-1).
\label{acesneu}
\end{equation}
Besides $M_{\tilde\nu}$, the cross section is proportional
to the $R$-parity violating product $X=(\lambda^\prime)^2B_l$
where $B_l$ is the sneutrino leptonic branching ratio
and $\lambda^\prime$ the relevant sneutrino
coupling to the ${\bar d} d$ quarks. Current constraints
on $X$ are very loose (we may take the range
$10^{-5}<X<10^{-1}$), and the 95\% CL Tevatron lower limits
on $M_{\tilde\nu}$ vary from 397 GeV ($X=10^{-4}$) to
866 GeV ($X=10^{-2}$) \cite{Aaltonen:2008ah}.

\section{RESULTS FOR SPIN-2 IDENTIFICATION}
There are wide domains in $M_R$ and
coupling constant values allowed by current
experimental limits where the scenarios described
above predict the same peaks in $M$
with the same numbers of events, so that they cannot be
distinguished from each other on the basis of the
event rates only.
\par
To proceed with the spin-identifying angular analysis, we suppose
that a peak at $M=M_R$ is discovered in process (\ref{proc}), and
assume that it is consistent with a spin-2 RS graviton (in which
case $M_R\equiv M_G$). To evaluate the domain in the ($M_G,c$)
plane, where the competitor spin-1 and spin-0 hypotheses with the same
number of events under the $M=M_G$ peak as the graviton hypothesis can
be excluded, hence the spin-2 hypothesis can be established, we
look at the `distances' among models:
\begin{equation}
\Delta A_{\rm CE}^{Z^\prime}=A_{\rm CE}^G-A_{\rm CE}^{Z^\prime}
\qquad {\rm and} \qquad
\Delta A_{\rm CE}^{\tilde\nu}=
A_{\rm CE}^G-A_{\rm CE}^{\tilde\nu}.
\label{deltaGSV}
\end{equation}
We adopt a simple-minded $\chi^2$-like criterion, where the
deviations (\ref{deltaGSV}) are compared to the expected
statistical uncertainty $\delta A_{\rm CE}^G$ pertinent to the RS
model (systematic uncertainties can also be included). The
$\chi^2$ for the $Z^\prime$ and the sneutrino cases in
(\ref{deltaGSV}) is defined as $\chi^2= \vert\Delta A_{\rm
CE}^{{Z^\prime},{\tilde\nu}}/ \delta A_{\rm CE}^G \vert^2$. The
desired {\it identification} domain in ($M_G,c$) is determined by
the condition $\chi^2\geq\chi^2_{\rm CL}$ for both the $Z^\prime$
and the $\tilde\nu$ models, with $\chi^2_{\rm CL}$ a critical
number that specifies the confidence level (3.84 for 95\% CL). Of
course, Eq.~(\ref{deltaGSV}) depends on the value of $z^*$, but it
turns out that the choice $z^*=0.5$ is `optimal' in maximizing the
sensitivity of the numerical analysis to the RS graviton.
\begin{figure}
  \includegraphics[width=.95\textwidth]{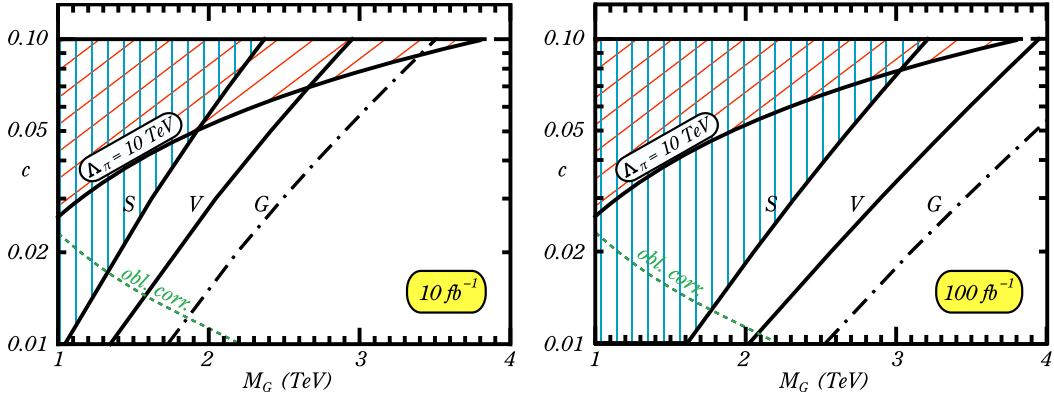}
  \caption{\label{fig1}Discovery and identification ranges
as defined in the text.}
\end{figure}

\par
Figure~\ref{fig1} shows the 5-$\sigma$ discovery domains and the
95\% identification domains for LHC luminosities of 10 and 100
${\rm fb}^{-1}$ ($l=e,\mu$ combined), and the `theoretically
favoured' restriction $\Lambda_\pi<10$ TeV taken into account.
Specifically: the area at the left of the line ``$G$'' is the
discovery domain for the lowest-lying RS graviton resonance; the
area at the left of the ``$V$'' curve is the exclusion domain of
the spin-1 hypothesis; finally, in the area at the left of the
``$S$'' line the spin-0 (and, as one can see, forcefully the
spin-1) hypothesis can be excluded. Therefore, the {\it
identification} domain where both spin-1 and spin-0 hypotheses can
be excluded and the spin-2 established is the intersection of the
``$S$'' area with the $\Lambda_\pi<10$~TeV one, i.e., above that 
curve. The allowed domain
to the right of the dashed ``oblique corrections'' line is
qualitatively determined by a fit to the oblique EW parameters
\cite{Davoudiasl:2000jd,Han:2000gp}, but the condition
dramatically restricting the discovery domains to the dashed
areas, if literally applied, is the $\Lambda_\pi$ bound. This
condition essentially forbids the creation of additional mass
scale hierarchies in the model.
Table~\ref{tab:a} summarizes the above results obtainable from the
observation of RS graviton peaks in dilepton final states.

\begin{table}
\begin{tabular}{lrrrr}
\hline
 & \tablehead{2}{r}{b}{Discovery}
 & \tablehead{2}{r}{b}{Identification} \\
\hline
${\cal L}_{\rm int}$ & $c=0.01$ &  $c=0.1$ &  $c=0.01$ &  $c=0.1$ \\
\hline
$10~{\rm fb}^{-1}$ & 1.7~TeV & 3.5~TeV & 1.1~TeV    & 2.4~TeV\\
$100~{\rm fb}^{-1}$ & 2.5~TeV & 4.6~TeV & 1.6~TeV  & 3.2~TeV\\
\hline
\end{tabular}
\caption{Discovery and Identification reach [TeV]} \label{tab:a}
\end{table}
\begin{figure}
  \includegraphics[width=.95\textwidth]{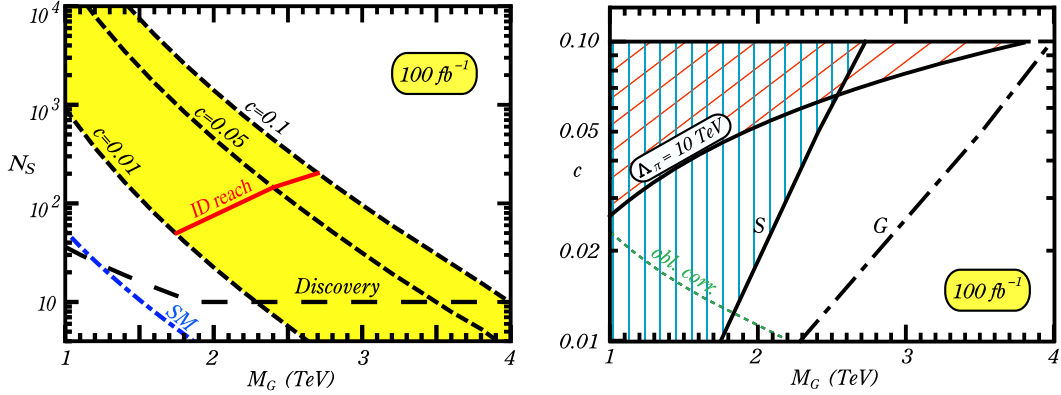}
  \caption{\label{fig2} Discovery and identification from the
process $pp \to G \to \gamma\gamma+X$}
\end{figure}

\par
Figure~\ref{fig2} shows some preliminary attempts to assess both
discovery and identification reaches on the spin-2 RS graviton
from the diphoton production channel in (\ref{proc}).
Specifically, the left panel shows, for $0.01<c<0.1$ the minimum
number of events needed to identify by means of the asymmetry
$A_{\rm CE}$ a peak at diphoton invariant mass $M_G$ as an RS
resonance, against the spin-0 hypothesis (95\% CL and LHC
luminosity of 100~${\rm fb}^{-1}$). Leading order QCD (therefore
unit $K$-factors) has been used to obtain this figure. Also, the
foreseen phase space experimental cuts for this channel have been
taken into account: $p_\bot >40$~GeV and
$\vert\eta\vert < 2.4$ for both photons,
and reconstruction efficiency 80\%.
The right panel of Fig.~\ref{fig2} represents, in the same style
as Fig.~\ref{fig1}, the translation to the ($M_G,c$) RS parameter
plane of the minimum number of diphoton events for RS graviton
identification given in the left panel. This channel is still
under study, in particular as regards the effect of the
next-to-leading order QCD effects and the perspectives at lower
LHC integrated luminosity. Nevertheless, the example in the figure
indicates that diphoton events may have a comparable
identification sensitivity to the spin-2 RS resonance, with the
spin-1 hypothesis automatically excluded.

\begin{theacknowledgments}
AP and AT acknowledge the partial support of the Belarusian
Republican Foundation for Fundamental Research. 
NP thanks the Organisers for the warm hospitality 
at a very stimulating and successful Workshop.
\end{theacknowledgments}

\bibliographystyle{aipproc}   

\end{document}